\documentclass[]{aa} 
\usepackage{amsmath}
\usepackage{graphicx}
\usepackage{graphics}
\usepackage{subfigure}
\usepackage{txfonts}
\usepackage{natbib}
\usepackage{xcolor}

\bibpunct{(}{)}{;}{a}{}{,}

\def\teff{\ifmmode T_{\rm eff} \else $T_{\mathrm{eff}}$\fi}

\def\ltsima{$\buildrel<\over\sim$}
\def\lsim{\lower.5ex\hbox{\ltsima}}

\newcommand{\ha}{\ifmmode {\rm H}\alpha \else H$\alpha$\fi}
\newcommand{\hb}{\ifmmode {\rm H}\beta \else H$\beta$\fi}
\newcommand{\lya}{\ifmmode {\rm Ly}\alpha \else Ly$\alpha$\fi}

\newcommand{\ebv}{\ifmmode E_{\rm B-V} \else $E_{\rm B-V}$\fi}
\newcommand{\av}{\ifmmode A_{\rm V} \else $A_{\rm V}$\fi}
\def\micron{$\mu$m}
\def\kms{km s$^{-1}$}

\def\msun{\ifmmode M_{\odot} \else M$_{\odot}$\fi}
\def\msunyr{\ifmmode M_{\odot} {\rm yr}^{-1} \else M$_{\odot}$ yr$^{-1}$\fi}
\def\zsun{\ifmmode Z_{\odot} \else Z$_{\odot}$\fi}

\def\lsun{\ifmmode L_{\odot} \else L$_{\odot}$\fi}

\def\mup{\ifmmode M_{\rm up} \else M$_{\rm up}$\fi}
\def\mlow{\ifmmode M_{\rm low} \else M$_{\rm low}$\fi}

\newcommand{\oh}{\ifmmode 12 + \log({\rm O/H}) \else$12 + \log({\rm
O/H})$\fi}

\newcommand{\cii}{[C~{\sc ii}]}

\def\Oiii{[O~{\sc iii}] $\lambda\lambda$4959,5007}

\def\flyf{\ifmmode f_{\rm Lyf} \else $f_{\rm Lyf}$\fi}
\def\pz{\ifmmode P(z) \else $P(z)$\fi}
\def\ki2{\ifmmode \chi^2 \else $\chi^2$\fi}
\def\zphot{\ifmmode z_{\rm phot} \else $z_{\rm phot}$\fi}

\newcommand{\xphot}{\ifmmode x_\gamma \else $v_\gamma$\fi}
\newcommand{\xobs}{\ifmmode x_{\rm obs} \else $x_{\rm obs}$\fi}
\newcommand{\xcmf}{\ifmmode x_{\rm CMF} \else $x_{\rm CMF}$\fi}
\newcommand{\vexp}{\ifmmode V_{\rm exp} \else $V_{\rm exp}$\fi}
\newcommand{\vmax}{\ifmmode V_{\rm max} \else $V_{\rm max}$\fi}
\newcommand{\nh}{\ifmmode N_{\rm HI} \else $N_{\rm HI}$\fi}
\newcommand{\dv}{\ifmmode \Delta v({\rm em-abs}) \else $\Delta v({\rm em}-{\rm abs})$\fi}

\def\fesc{\ifmmode f_{\rm esc} \else $f_{\rm esc}$\fi}

\def\frellya{\ifmmode f^{\rm rel}_{\rm{Ly}\alpha} \else $f^{\rm rel}_{\rm{Ly}\alpha}$\fi}

\newcommand{\mstar}{\ifmmode M_\star \else $M_\star$\fi}
\newcommand{\mdust}{\ifmmode M_d \else $M_d$\fi}
\newcommand{\muv}{\ifmmode M_{1500} \else $M_{1500}$\fi}
\newcommand{\luv}{\ifmmode L_{\rm UV} \else $L_{\rm UV}$\fi}
\newcommand{\lir}{\ifmmode L_{\rm IR} \else $L_{\rm IR}$\fi}
\newcommand{\lbol}{\ifmmode L_{\rm bol} \else $L_{\rm bol}$\fi}
\newcommand{\liruv}{\ifmmode L_{\rm IR+UV} \else $L_{\rm IR+UV}$\fi}
\newcommand{\liroveruv}{\ifmmode L_{\rm IR}/L_{\rm UV} \else $L_{\rm IR}/L_{\rm UV}$\fi}
\newcommand{\nlyc}{\ifmmode N_{\rm Lyc} \else $N_{\rm Lyc} $\fi}
\newcommand{\rholyc}{\ifmmode \rho_{\rm Lyc} \else $\rho_{\rm Lyc} $\fi}
\newcommand{\auv}{\ifmmode  A_{\rm UV} \else $A_{\rm UV}$\fi}

\newcommand{\lcii}{\ifmmode L_{[\rm CII]} \else $L_{[\rm CII]}$\fi}
\newcommand{\lirngc}{\ifmmode L_{\rm IR}^{\rm N6946} \else $L_{\rm IR}^{\rm N6946}$\fi}

\newcommand{\abell}{A1703-zD1}
\newcommand{\finkel}{z8-GND-5296}

\begin{document}
    \title{New constraints on dust emission and UV attenuation of
$z=6.5-7.5$ galaxies from millimeter observations\thanks{Based on observations carried out with the IRAM Plateau de Bure 
Interferometer and the IRAM 30 m telescope. IRAM is supported by CNRS/INSU (France), the MPG (Germany) and the IGN (Spain).}}  
\subtitle{}
  \author{D. Schaerer\inst{1,2}, F. Boone\inst{2}, M. Zamojski\inst{1}, J. Staguhn\inst{3,4}, M. Dessauges-Zavadsky\inst{1}, S. Finkelstein\inst{5}, 
  F. Combes\inst{6},}
  \institute{
Observatoire de Gen\`eve, Universit\'e de Gen\`eve, 51 Ch. des Maillettes, 1290 Versoix, Switzerland
         \and
CNRS, IRAP, 14 Avenue E. Belin, 31400 Toulouse, France
	\and	
	The Henry A. Rowland Department of Physics and Astronomy, Johns Hopkins University, 3400 N. Charles Street, Baltimore, MD 21218, USA \and
Observational Cosmology Lab., Code 665, NASA at Goddard Space Flight Center, Greenbelt, MD 20771, USA \and
The University of Texas at Austin, Austin, TX 78712, USA \and	
Observatoire de Paris, LERMA, 61 Av. de l'Observatoire, 75014, Paris, France 
}

\authorrunning{D. Schaerer et al.}
\titlerunning{Dust emission and UV attenuation of $z \sim 6.5-7.5$ galaxies from millimeter observations}

\date{Received date; accepted date}

\abstract{Determining the dust properties and UV attenuation of distant star-forming galaxies is of great interest
for our understanding of galaxy formation and cosmic star formation in the early Universe. However, few 
direct measurements exist so far.}
{To shed new light on these questions we have targeted two recently discovered Lyman break galaxies (LBGs) at $z \approx 6.8$ and 
$z=7.508$  to search for dust continuum and  \cii\ $\lambda$158\micron\ line emission.}
{The strongly lensed $z \approx 6.8$ LBG \abell\ behind the galaxy cluster Abell 1703, and the spectroscopically 
confirmed $z=7.508$ LBG \finkel\ in the GOODS-N field have been observed with the Plateau de Bure interferometer (PdBI) 
at 1.2mm. These observations have been combined with those of three $z>6.5$ \lya\ emitters (named HCM6A, Himiko, and IOK-1),
for which deep measurements were recently obtained with the PdBI and ALMA.}
{\cii\ is undetected in both galaxies, providing a deep upper limit of $\lcii < 2.8 \times 10^7$ \lsun\ for \abell,
comparable to the non-detections of Himiko and IOK-1 with ALMA.
Dust continuum emission from \abell\ and \finkel\ is not detected with an rms of 0.12 and 0.16 mJy/beam. 
From these non-detections and earlier multi-wavelength 
observations we derive upper limits on their IR luminosity and star formation rate, dust mass, and UV attenuation.
Thanks to strong gravitational lensing the achieved limit for \abell\ is comparable to those achieved with ALMA,
probing the sub-LIRG regime ($\lir<8.1 \times 10^{10}$ \lsun) and very low dust masses ($M_d<1.6 \times 10^7$ \msun).
We find that all five galaxies are compatible with the Calzetti IRX--$\beta$ relation, 
their UV attenuation is compatible with several indirect estimates from other methods (the UV slope, 
extrapolation of the attenuation measured from the IR/UV ratio at lower redshift, and SED fits),
and the dust-to-stellar mass ratio is not incompatible with that of galaxies from $z=0$ to 3. 
For their stellar mass the high-$z$ galaxies studied here have an attenuation below the one expected from the 
mean relation of low redshift ($z\la 1.5$) galaxies.
}
{More and deeper (sub)-mm data are clearly needed to directly determine the UV attenuation and dust content 
of the dominant population of high-$z$ star-forming galaxies and to establish more firmly their dependence
on stellar mass, redshift, and other properties.}


  \maketitle

\section{Introduction}
\label{s_intro}

Both the strong interest for galaxy formation and evolution in the early Universe and the high sensitivity of 
millimeter interferometers meet ideally at high redshifts. Indeed, due to the well-known so-called 
negative k-correction, the observed flux from dust emission in galaxies {\em increases} with increasing
distance at (sub)millimeter wavelengths, allowing thus in principle observations out to very high redshifts, 
$z>6$ \citep[cf.][]{Blain2002Submillimeter-g}. 
Since galaxy evolution in the early Universe remains poorly understood and since dust can strongly affect 
our view of star formation, it is essential to observe directly dust emission from the most distant galaxies.

The Lyman break selection has been successful in tracing a large number of galaxies
out to very high redshift, which has allowed the construction of their luminosity function and thus 
determinations of the cosmic UV luminosity density and the associated star formation rate density 
\citep[][and references therein]{Madau2014Cosmic-Star-For}. Despite this, little is known about the UV attenuation
of these galaxies and about the associated dust properties (mass, temperature, composition etc.).
Generally, the UV attenuation of Lyman break galaxies (LBGs) at $z >3$ is estimated from their UV slope $\beta$ 
\citep[e.g.][]{Bouwens2013,castellanoetal2012}
using empirical relations established at low redshift \citep[e.g.\ the so-called ``Meurer relation", cf.][]{meurer99}, or from
SED fits to the (rest frame) UV--optical part of the spectrum. 
Although the most direct measure of UV attenuation is determined by the ratio of IR/UV luminosity
\citep{Buat05,igleasias-paramo2007}, current measurements are limited by sensitivity to $z \la 3-4$
\citep{Lee2012Herschel-Detect,burgarella2013}.

Thanks to strong gravitational lensing, the normal sensitivity limit has been overcome in
few cases in the past. E.g.\ \cite{Livermore2012Observational-l} has achieved a tentative detection
of CO in a strongly lensed $z=4.9$ star-forming galaxy with a star formation rate SFR $\approx 40$ \msunyr\ 
and stellar mass $\mstar \approx 7.\times 10^8$ \msun, i.e.\ properties comparable of LBGs.
The IR continuum of this galaxy has, however, remained undetected, with a limit corresponding
to $\lir < 3.5 \times 10^{11}$ \lsun, as reported by \cite{Livermore2012Observational-l}.
The $z=6.56$ \lya\ emitting galaxy (LAE) HCM6A, lensed by the galaxy cluster Abell 370, has been
targeted by various studies aiming at detecting the dust continuum, CO, and \cii\ emission
\citep{Boone2007Millimeter-obse,2009ApJ...697L..33W,2013ApJ...771L..20K}. 
From the non-detection at 1.2mm, obtained with MAMBO-2 at the IRAM 30m antenna,
\cite{Boone2007Millimeter-obse}, derived an upper limit of $\lir/\luv \la 2$ and an UV attenuation of  $\auv \la 0.9$, 
which we here update using a more recent measurement from \cite{2013ApJ...771L..20K}.
Until recently, this measurement has represented the best (lowest) constraint on UV attenuation in
LBGs or LAE at high redshift ($z>5$).

Two bright non-lensed LAEs, Himiko at $z=6.595$ and IOK-1 at $z=6.96$ \citep{2009ApJ...696.1164O,Iye06},
and the gamma-ray burst 090423 at $z \sim 8.2$ were also targeted by \cite{Walter2012Evidence-for-Lo} with the
Plateau de Bure interferometer (PdBI) in the 1.2mm window, from which they conclude  
a low dust obscuration in these objects.
Two other LAEs at $z \sim 6.5$, recently observed by \cite{Gonzalez-Lopez2014Search-for-C-II} with the PdBI in the 1.2mm window,
were also not detected (both in \cii\ and in the continuum). However, since their continuum sensitivity is a factor 2--5 lower than 
the one achieved for the lensed galaxy HCM6A and this galaxy is magnified by a factor $\mu \approx 4.5$, 
the effective limit on the \lir/\luv\ ratio, and hence the UV attenuation of these galaxies, is significantly less stringent
than the limit obtained for HCM6A.

Himiko and IOK-1 have recently been observed with ALMA \citep{Ouchi2013An-Intensely-St,Ota2014ALMA-Observatio}
reaching unprecedented depths in the 1.2mm window (band 6), a factor 6--10 times more sensitive than the PdBI
observations of HCM6A of  \cite{2013ApJ...771L..20K}.
Here, we combine these deep measurements with new PdBI observations of two interesting, bright high-$z$ LBGs,
the strongly lensed galaxy \abell\ with a well-defined photometric redshift of $z \approx 6.8$ discovered by 
\cite{Bradley2012Through-the-Loo},
and \finkel, one of the most distant spectroscopically confirmed galaxies with $z=7.508$ \citep{2013Natur.502..524F}.
Indeed, with an apparent H-band magnitude of 24, boosted by gravitational lensing by a factor $\sim$ 9
\citep{Bradley2012Through-the-Loo}, \abell\ is one of the brightest $z \sim 7$ galaxies, outshining 
Himiko and IOK-1 by $\sim$ 1--1.6 magnitude in the near-IR (UV rest frame).
The analysis of \finkel\ yielding indications for non-negligible UV attenuation and possibly an SFR in
excess of 300 \msunyr\ \citep{2013Natur.502..524F}, these two galaxies are obviously interesting
targets for deep (sub)-mm observations.

Tuning the PdBI to the frequency of the \cii\ line, we have obtained  observations  reaching a continuum sensitivity 
of 0.12-0.16 mJy in the 1.2mm window for these two high-z objects. The resulting continuum non-detections
are used to provide limits on their IR luminosity, dust mass, and UV attenuation. Together with this
data we  consistently analyze the deepest available IR/mm data of Himiko, IOK-1, and HCM6A,
confronting in particular the derived UV attenuation with expectations for high redshift LBGs,  
discussing the attenuation and dust mass as a function of stellar mass, and how these quantities
may evolve with redshift. 

Our paper is structured as follows. The observational data are described in Sect.\ \ref{s_obs}. In Sect.\ \ref{s_ir} we derive 
the IR luminosity and dust masses from the observations. Simple inferences on the UV attenuation and the stellar populations
of the small sample are derived in Sect.\ \ref{s_stellar}. We then discuss our results concerning the IRX--$\beta$ relation,
UV attenuation as a function of redshift and stellar mass,  dust mass, the \cii\ luminosity, and constraints
on the SED fits of these objects in Sect.\ \ref{s_discuss}.
Section \ref{s_conclude} summarizes our main conclusions.
We adopt a $\Lambda$-CDM cosmological model with $H_{0}$=70 km s$^{-1}$ Mpc$^{-1}$, 
$\Omega_{m}$=0.3 and $\Omega_{\Lambda}$=0.7, use AB magnitudes, and assume a Salpeter IMF from 0.1 to 100 \msun.

\section{Observations}
\label{s_obs}

\subsection{IRAM observations of the lensed galaxy \abell\ and the GOODS-N source \finkel}
We have obtained 1.2 mm observations of two sources,   \abell\ and  \finkel, with the 
IRAM PdBI in the compact configuration and with the WIDEX
backend covering a bandwidth of 3.6\,GHz. \abell\ was observed in
May and August 2013 with 5 antennas and \finkel\ was observed in
December 2013 with 6 antennas. The central frequency of the receiver 3
was tuned to the redshifted \cii\ line, whose rest frequency is
1900.54 GHz, i.e.,  241.5\,GHz for A1703-zD1  and 223.88\,GHz for
z8-GND-5296. The QSOs, 1150+497, J1259+516 and 1347+539 were used as
phase and amplitude calibrators for the observations of \abell\ and
1150+497, and 1044+719 for the observations of \finkel. The bright
sources 2200+420, MWC349, 3C84 and LKHA101 were also observed for the
absolute flux and passband calibration.

The Widex correlator covers the frequency range 240.2-243.8\,GHz and 222.082-225.682\,GHz, 
corresponding to the redshift ranges 6.796-6.912 and 7.422-7.557
for \abell\ and \finkel\ respectively.
The redshift of \abell\ is estimated from photometric observations. 
\cite{Bradley2012Through-the-Loo} find $\zphot=6.7^{+0.2}_{-0.1}$,  \cite{2014ApJ...784...58S} $\zphot=6.8\pm0.1$,
whereas we obtain a 68\% confidence range of 6.7--6.9, with a median between 6.79 and 6.86, depending on which 
IRAC photometry is used (cf.\ below).  Although \zphot\ is fairly well defined,
the exact probability to cover the \cii\ line is difficult to assess.
For z8-GND, however, the redshift is based on the detection of the \lya\ line and the band 
covered corresponds to a velocity range $[$-3078,1752$]$ \kms\ with respect to the systemic redshift
determined from \lya. Our tuning should therefore fully cover \cii\ emission from this galaxy.

The data reduction was done using the GILDAS software. The data cubes
were build using natural weighting to maximize the point source
sensitivity. The beam sizes obtained are: 2.30$''$$\times$1.75$''$
and 1.78$''$$\times$1.61$''$ for \abell\ and \finkel\
respectively. The continuum rms are 165 and 124 $\mu$Jy and the line
rms in 50\,km\,s$^{-1}$ channels are 1.52 and 1.83 mJy, respectively.
None of the sources is detected.
In passing we note also that none of the three other, near-IR fainter, z-drop galaxies (A1703-zD3, A1703-zD6, and A1703-zD7) 
from \cite{Bradley2012Through-the-Loo}, which are also included in our field, is detected.

In an earlier observing run with the IRAM 30m telescope, we have also obtained a deep
2mm map of the Abell 1703 cluster with GISMO. No source was detected on the image
reaching a depth of 0.15 mJy rms/beam.
We have also examined the available Herschel PACS and SPIRE images of this cluster
taking within the Herschel Lensing Survey \citep{Egami2010The-Herschel-Le}.
As expected, the galaxy \abell\ is not detected down to $\sim$ 1 mJy at 500 \micron,
and 40 mJy at the shortest wavelengths (70 \micron).
In any case, as the most constraining data comes from the 1.2mm PdBI flux, the
Herschel and GISMO limits are not used here to determine limits on the IR luminosity and dust mass.
 
\begin{table*}[htb]
\caption{Summary of millimeter observations and derived quantities. All luminosity upper limits are 3 $\sigma$ and are {\em not} corrected for lensing. 
For \abell\ and HCM6A the true luminosity limits are therefore lower by the magnification factor $\mu$. 
The dust temperature $T_d$ indicated here is corrected for the CMB heating, i.e., it corresponds to the temperature dust 
would have if it were heated by stars only.}
\begin{tabular}{lllllllllllllll}
\hline
\hline
Source & $z$ & $\nu$  & rms$_{\rm cont}$  &  $\sigma_{\rm line}$  & \lcii\  & $\lir(T_d=25)$ & $\lir(T_d=35)$  & $\lir(T_d=45)$  & $\mu$ \\
           &        & [GHz] & [mJy beam$^{-1}$] & [mJy beam$^{-1}$]$^e$ & $10^8$ [\lsun] &  $10^{11}$ [\lsun]  & $10^{11}$ [\lsun]  & $10^{11}$ [\lsun] \\
\hline
\abell\ & 6.8$^a$    & 241.500 & 0.165 & 1.517 & $<2.55/\mu$ & $<3.96/\mu$ & $<7.32/\mu$ & $<14.38/\mu$ & 9. \\
\finkel\ & 7.508 & 223.382 & 0.124 & 1.824 & $<3.56$ & $<3.84$ & $<6.65$ & $<12.67$ \\
\\
IOK-1$^b$    & 6.96 & 238.76   & 0.021 & 0.215 & $<0.38$ &  $<0.53$ & $<0.96$ & $<1.87$ \\ 
HCM6A$^c$ & 6.56 & 251.40   & 0.16   & 0.849 & $<1.36/\mu$ &  $<3.47/\mu$ & $<6.49/\mu$ & $<12.81/\mu$ & 4.5 \\
Himiko$^d$ & 6.595 & 250.00 & 0.017 & 0.167 & $<0.28$ & $<0.36$ & $<0.67$ & $<1.30$ \\
%
\hline  
\multicolumn{9}{l}{$^a$ Approximate photometric redshift (cf.\ text). 
			$^b$ Observations from \citet{Ota2014ALMA-Observatio}.
			$^c$ Observations from \citet{2013ApJ...771L..20K}.} \\
\multicolumn{9}{l}{$^d$ Observations from \citet{Ouchi2013An-Intensely-St}. $^e$ In $\Delta v=50$ \kms\ channels.}\\
\label{t_ir}
\end{tabular}
\end{table*}

\subsection{Other data}
\label{s_otherobs}
From the literature we compiled the visible to near-IR (8 \micron) data for \abell\ and \finkel.
The HST and IRAC photometry for \abell\ is taken from \citet{Bradley2012Through-the-Loo}. 
\citet{2014ApJ...784...58S} have remeasured the photometry of this object, finding 
differences in the IRAC filters ($m_{3.6}=23.66$ and $m_{4.5}=24.93$, Smit 2014, private communication),
which translates to a larger 3.6 \micron\ excess than the data of \citet{Bradley2012Through-the-Loo}. We have therefore modeled both sets of photometry.
The photometry of \finkel\ is taken from \cite{2013Natur.502..524F}.

We also analyze three other related $z>6$ objects for comparison: the strongly lensed $z=6.56$ \lya\ emitter
HCM6A, the $z=6.96$ \lya\ emitter IOK-1, and the bright $z=6.595$ \lya\ blob called Himiko, which 
were previously observed at (sub)millimeter wavelengths with IRAM and with ALMA 
\citep{Boone2007Millimeter-obse,Walter2012Evidence-for-Lo,2013ApJ...771L..20K,Ouchi2013An-Intensely-St,Ota2014ALMA-Observatio}.
For HCM6A we use the recent IRAM data from \citet{2013ApJ...771L..20K}, which are somewhat deeper 
than our earlier MAMBO-2 observations.
The ALMA observations of IOK-1 and Himiko are described in detail in \citet{Ota2014ALMA-Observatio}
and \citet{Ouchi2013An-Intensely-St}.
The corresponding millimeter observations (also non-detections) are summarized in Table~\ref{t_ir}.

All three objects have photometry in the near-IR (HST plus ground-based) and in the IRAC bands.
Photometry for HCM6A has been compiled in \citet{Boone2007Millimeter-obse}; 
\citet{2011ApJ...735L..38C}
have obtained more recent measurements with WFC3/HST. 
The IRAC photometry of this galaxy is difficult/inconsistent due to contamination by neighboring sources. 
We therefore refrain from presenting detailed updated SED fits for this object
\citep[cf.][]{SP05,2005ApJ...635L...5C,2013ApJ...771L..20K}.
%
For IOK-1 we use the WFC3/HST photometry of \citet{2011ApJ...736L..28C}
and the IRAC data from Egami (2014, private communication).
The total magnitudes for Himiko are taken from \citet{Ouchi2013An-Intensely-St}.


Other $z>6$ LBGs and LAEs have recently been observed in the mm-domain, but are not included
in our comparison, since the limits on their dust mass and UV attenuation are significantly less stringent
than the ones for the objects listed in Table~\ref{t_ir}.
This is the case for two other LAEs with confirmed spectroscopic redshifts at $z \sim 6.5$ that were recently observed
at 1.2mm with CARMA to search for \cii\ emission, and remained also undetected in the continuum \citep{Gonzalez-Lopez2014Search-for-C-II}.
Their observations are factor 2--5 less deep than those of \cite{2013ApJ...771L..20K} for HCM6A, which furthermore is magnified by 
a factor $\sim 4.5$. Although their UV magnitudes are comparable to the intrinsic, i.e.\ lensing-corrected, one of HCM6A
the constraint on \lir/\luv, hence UV attenuation, is therefore clearly less strong than for HCM6A.
We also chose not to include the $z \sim 9.6$ lensed galaxy candidate  of \citet{2012Natur.489..406Z} recently
discussed by \cite{Dwek2014Dust-Formation-},  since its association with the MACS1149-JD source is still inconclusive.

\subsection{Observed SEDs}

The ``global" SEDs of \abell\ and \finkel\ from the near-IR to the millimeter domain are found to be similar
to those of the other objects also included here, which are HCM6A, IOK-1, and Himiko, and are therefore 
not shown here. Schematically, they are characterized by a relatively low IR/mm emission with respect to
their rest-frame optical emission, comparable to local dwarf galaxies and excluding SEDs of local ULIRGs 
or dusty star-forming galaxies such as Arp 220 or M82, or even more normal spiral
galaxies as NGC 6949 \citep[see][]{Boone2007Millimeter-obse,Walter2012Evidence-for-Lo,Ouchi2013An-Intensely-St,Gonzalez-Lopez2014Search-for-C-II,Ota2014ALMA-Observatio,Riechers2014ALMA-Imaging-of}.

\section{IR and dust properties}
\label{s_ir}

The 1.2mm observations listed in Table \ref{t_ir} were used to determine limits on the \cii\ 158 \micron\ luminosity, \lcii,
the total IR luminosity, \lir, and the dust mass, $M_d$. The results are given in Table \ref{t_ir} and \ref{t_mdust}
for three different values of the dust temperature $T_d$.

The upper limits on the \cii\ line luminosities are computed by assuming a line width $\Delta v=50$\,km\,s$^{-1}$ to be 
consistent with \cite{Gonzalez-Lopez2014Search-for-C-II} and by applying $L_{\rm CII}=1.04\times10^{-3} S_{\rm CII} \Delta v (1+z)^{-1}D^2_L$ 
(Solomon et al 1992), where $S_{\rm CII}$ is the line flux and $D_L$ the luminosity distance. 
Assuming a narrow line width results in a conservative estimate of the upper limit.
After correction for lensing the upper limits on the \cii\ luminosity are very similar for all galaxies,
$\log(\lcii) < 7.45-7.6$ \lsun, except for \finkel, where the upper limit is approximately a
factor 10 higher.

We compute the mass of dust by assuming a dust mass absorption
coefficient $\kappa_{\nu}=1.875(\nu/\nu_0)^{\beta_{\rm IR}}$\,m$^2$kg$^{-1}$ with
$\nu_0=239.84$\,GHz and $\beta_{\rm IR}=1.5$ and by removing the
contribution of the CMB to the dust heating as detailed by \citet{2013ApJ...766...13D}
and in a similar way to \citet{Ota2014ALMA-Observatio}. We also compute
the IR luminosity of the dust heated by the stars (CMB contribution removed) 
by integrating the SED between 8 and  1000\micron\
assuming a modified black body  SED with a power law in the Wien
regime with a spectral index $\alpha=2.9.$

To examine  how \lir\ and \mdust\ depend on the unknown dust temperature, we have assumed 
three different values $T_d=25$, 35, and 45 K, where $T_d$ is the dust temperature prior to correction
for the CMB (i.e.\ $T_d$ is the temperature the dust would have without the CMB heating).
Typically the IR luminosity changes by a factor $\la 3.2-3.7$ for the range of dust temperatures ($T_d=25$ to 45 K), translating
to an uncertainty of $\approx \pm 0.25$ dex.
At $z \sim 7$ the CMB temperature is $\sim 22$ K. 
\citet{Ota2014ALMA-Observatio} find $T_d=27.6$ K for the average dust temperature of local dwarf and irregular galaxies, whose SED
may be comparable to that of the high-$z$ sources (cf.\ above).
Slightly higher temperatures, $T_d \sim 30-35$~K are  found by \cite{Hirashita2014Constraining-du} for Himiko from dust modeling.
From empirical arguments  a higher dust temperature might be more appropriate for the high-z galaxies
\cite[cf.][]{2014A&A...561A..86M,Sklias2014Star-formation-}.
Dust masses change approximately by a factor $\sim 7$ for $T_d=25$ to 45 K.
For simplicity we will subsequently adopt \lir\ and \mdust\ values derived for the intermediate value of $T_d=35$ K.

\begin{table}[htb]
\caption{Derived dust mass limits assuming different dust temperatures 
$T_{\rm d}$, a magnification $\mu=4.5$, a dust mass absorption coefficient $\kappa_{125}=1.875$ m$^2$ kg$^{-1}$, 
and a dust emissivity index $\beta_{\rm IR}=1.5$.
All limits are $3 \sigma$ upper limits and are {\em not} corrected for lensing. 
For \abell\ and HCM6A the true dust mass limits are therefore lower by a factor $\mu=9.$ and 4.5 respectively.}
\begin{tabular}{lrllll}
\hline
\hline
Source             & $M_d(T_{\rm d}=25)$             & $M_d(T_{\rm d}=35)$           & $M_d(T_{\rm d}=45)$ \\
                        & $10^7$ \msun & $10^7$ \msun & $10^7$ \msun \\
\hline
\abell  & $<51.46/\mu$  & $<14.95/\mu$ & $<7.41/\mu$  \\
\finkel & $<49.88$  & $<13.59$  & $<6.53$          \\
\\
IOK-1  & $<6.87$ & $<1.96$ & $<0.96$ \\
HCM6A & $<45.00/\mu$ & $<13.26/\mu$ & $<6.60/\mu$ \\
Himiko & $<4.72$ &  $<1.36$ & $<0.67$ \\
\hline
\label{t_mdust}
\end{tabular}
\end{table}

\section{Stellar populations and the UV attenuation}
\label{s_stellar}

\subsection{Observed and derived properties}
\label{s_beta}
The absolute UV magnitude \muv\ and UV luminosity, UV star formation rate (neglecting dust attenuation
and assuming the \citet{Kennicutt1998} calibration), UV slope $\beta$  defined as $F_\lambda \propto \lambda^ \beta$ in a standard manner, 
the UV attenuation \auv, and 
the dust-obscured SFR(IR) of the objects are listed in Table \ref{t_derived}.

The UV slopes are derived from the UV photometry using two bands or from our SED fits
for these galaxies following well-known standard methods
\citep[e.g.][]{finkelsteinetal2011,Bouwens2013,Ouchi2013An-Intensely-St}.
The typical uncertainty and differences between these methods is 
$\pm( 0.3-0.5)$ on $\beta$, as also found comparing our values with those listed in the literature.
For \abell, e.g., \cite{2014ApJ...784...58S} find $\beta=-1.4 \pm 0.3$, whereas our SED fits yield
$\beta \approx -1.6$. For \finkel, using two WFC3 photometric bands, one obtains $\beta \approx-1.4$,
compared to  $\beta \approx -1.7$ from our SED fits. \cite{Ouchi2013An-Intensely-St} find
$\beta=-2.00 \pm 0.57$ for Himiko.

The UV attenuation is constrained by the observed limit on \lir/\luv.
We use the expression of  \citet{schaerer2013}  relating the UV attenuation factor at 1800 \AA, $f_{\rm UV}$, 
to $x=\log(\lir/\luv)$:
\begin{equation}
\log(f_{\rm UV}) = 0.24 + 0.44 x + 0.16 x^2.
\label{eq_auv}
\end{equation}
By definition one has $\auv = 2.5 \log f_{\rm UV}$. In Table \ref{t_derived} we list the upper limits on the UV attenuation
using the 3 $\sigma$ limits on $\lir(T_d=35)$, and $T_d=25$ and 45 K to estimate the uncertainty.
For a specific attenuation law this can of course be translated into
quantities such as \av. For the Calzetti law \citep{calzettietal2000}, e.g., $A_V = 2.5 (R_V/k_\lambda) \log f_{\rm UV}= 1.08 \log f_{\rm UV}$ = $0.43 \auv$,
and $E(B-V)=0.26 \log f_{\rm UV} = 0.11 \auv$.

\subsection{Stellar masses and star formation rates}
To obtain simple estimates of the SFR and stellar mass we use the ``classical" SFR(UV) calibration of \cite{Kennicutt1998}  
and the mean relation between the UV magnitude and stellar mass, 
\begin{equation}
\log(\mstar/\msun)=-0.45 \times (\muv+20) + 9.11,
\end{equation}
obtained by \cite{schaerer2014} from detailed SED
fits including nebular emission to a sample of LBGs (z-drops at $z \approx 6.7$). The corresponding values of SFR(UV) 
and $\mstar({\rm UV})$ are listed in Table \ref{t_derived}. The uncertainty of a factor $\sim 2$ in stellar mass corresponds 
to the differences obtained for different SF histories 
(exponentially rising histories yield the lowest masses, exponentially declining ones intermediate masses, and constant SFR 
the highest mass), as described in  \cite{schaerer2014}. 
In the last column of Table \ref{t_derived} we also list the upper limit of SFR(IR) derived from our limit on $\lir(T_d=35)$,
assuming again the ``standard" calibration of \cite{Kennicutt1998}.
Note that both SFR(UV) and SFR(IR) can underestimate the true star formation rate if the  population dominating the
UV emission is relatively young ($<100$ Myr), as SED fits for some galaxies of our sample indicate
\citep[cf.][]{2013Natur.502..524F}. For this reason, and since the UV star formation rate is not corrected for dust attenuation in
Table \ref{t_derived}, SFR(UV) represents most likely a true {\em lower} limit on the current SFR of these galaxies.

Overall the above simple estimates yield stellar masses between $\sim 10^9$ and $10^{10}$ \msun\ for the five galaxies
studied here. Generally speaking these values agree with earlier estimates, and the differences do not affect our
conclusions.
For \abell\ \cite{Bradley2012Through-the-Loo} find masses  in the range of $\mstar = (0.7-1.5) \times 10^9$ \msun, for a simple stellar population
and constant SFR; the higher value is in perfect agreement with our mass estimate. Their SFR $=7.3\pm0.3$ \msunyr\ is only slightly
lower than our SFR(UV) value, probably due to the sub-solar metallicity they adopt.
For IOK-1 \citet{2010PASJ...62.1167O} provide a rough mass estimate (in agreement with ours) spanning more than 1 dex
depending on the age of the dominant population. In deeper Spitzer observations, which are now available, this galaxy is detected
at 3.6 and 4.5 \micron, yielding a stellar mass of the order of $\sim 1.\ \times 10^{10}$ \msun\ (Egami et al., in preparation),
compatible with our estimate. 
Our mass estimate for HCM6A is a factor $\sim 2.4$  larger than that of \citet{2013ApJ...771L..20K}. \cite{SP05} infer a maximum mass 
of $\la 4 \times 10^9$ \msun\ after adjustment to the IMF assumed here.
Himiko is the most massive galaxy according to our estimates. \citet{Ouchi2013An-Intensely-St} determine $\mstar \approx 1.5 \times 10^{10}$ \msun,
a factor $\sim 2.3$ higher than our estimate. Their SFR$=100 \pm 2$ \msunyr\ from SED fits is most likely overestimated, as
it includes an attenuation corresponding to $A_V=0.6$, which is clearly ruled out by their limit on the dust attenuation.
Only for  \finkel\ our simple mass and SFR estimates differ from those derived previously. Indeed, for this galaxy the SEDs fits
favor very young ages (1-3 Myr), leading to a lower mass (by a factor $\sim 5$) and higher SFR (of the order of 300-1000 \msunyr),
according to \cite{2013Natur.502..524F}. 
In any case, our main conclusions regarding the mass dependence of dust attenuation and the dust mass are
not affected by these uncertainties on the stellar mass, as will be clear below (Sects.\ \ref{s_uv} and \ref{s_dust}).

\subsection{SED modelling}
To examine the constraints provided by the limits on UV attenuation on derived physical parameters
of our galaxies, we have also carried out SED fits of these objects. The code and the ingredients 
are the same as described in \cite{schaerer&debarros2009,schaerer&debarros2010} and \cite{schaerer2013}.
Basically, we use a large set of spectral templates based on
the GALAXEV synthesis models of \cite{bruzual&charlot2003}, covering
different metallicities and three different star formation histories. Nebular emission is added. 
A Salpeter IMF from 0.1 to 100 \msun\ is adopted. 
The free parameters of our SED fits are:
age $t_\star$ defined since the onset of star-formation, 
attenuation $A_V$ described by the Calzetti law \citep{calzettietal2000},
and metallicity $Z$ (of stars and gas).
For all but one object, the redshift is fixed to the value from spectroscopy; for \abell\ we
treat $z$ as a free parameter, although it is very well constrained by the sharp Lyman break \citep[cf.][]{Bradley2012Through-the-Loo}.

\begin{table*}[htb]
\caption{Derived quantities. For \abell\ we have assumed a magnification factor of $\mu=9$, for HCM6A $\mu=4.5$.
The upper limits on the UV attenuation and SFR(IR) are derived from $\lir(T_d=35)$ and using Eq.\ \ref{eq_auv}.
The SFR values (limit) listed here assume constant SF, ages $\protect\ga 100$ Myr, and the  \cite{Kennicutt1998} calibration.
SFR(UV) is not corrected for reddening. Typical uncertainties for the UV slope $\beta$ are $\pm (0.3-0.5)$ (cf.\ Sect.\ \ref{s_beta}).}
\begin{tabular}{llllllllllllll}
\hline
\hline
Source &  \muv & $\log(\luv)$ & SFR(UV) & $\log(\mstar({\rm UV})) $ &  $\beta$ & \auv & SFR(IR)  \\
           &  [mag]   & [\lsun] & [\msunyr] & [\msun] & & [mag] & [\msunyr] \\
\hline
\abell\  & -20.3  & 10.45 & 9.0 &    $9.2\pm 0.3$ & -1.4  &  $<1.2\pm 0.4$  & $<13.8$  \\
\finkel\ & -21.4  &  10.86 & 23.4 & $9.7\pm 0.3$  & -1.4             & $<2.0^{+0.5}_{-0.4}$     & $<113$  \\
IOK-1  & -21.3  & 10.8 & 20.4 &    $9.7\pm 0.3$   & -2.0                & $<0.8^{+0.5}_{-0.3}$     & $<16.3$  \\ 
HCM6A &  -20.8 & 10.63 & 13.7 & $9.5\pm 0.3$   & -1.7           & $<1.3\pm 0.4$          & $<24.5$  \\
Himiko & -21.7 & 11.0  & 32.3 &  $9.8\pm 0.3$   &-2.0 &   $<0.4^{+0.3}_{-0.2}$ & $<11.4$ \\
\hline  
\label{t_derived}
\end{tabular}
\end{table*}

\begin{figure}[htb]
\centering
\includegraphics[width=8.8cm]{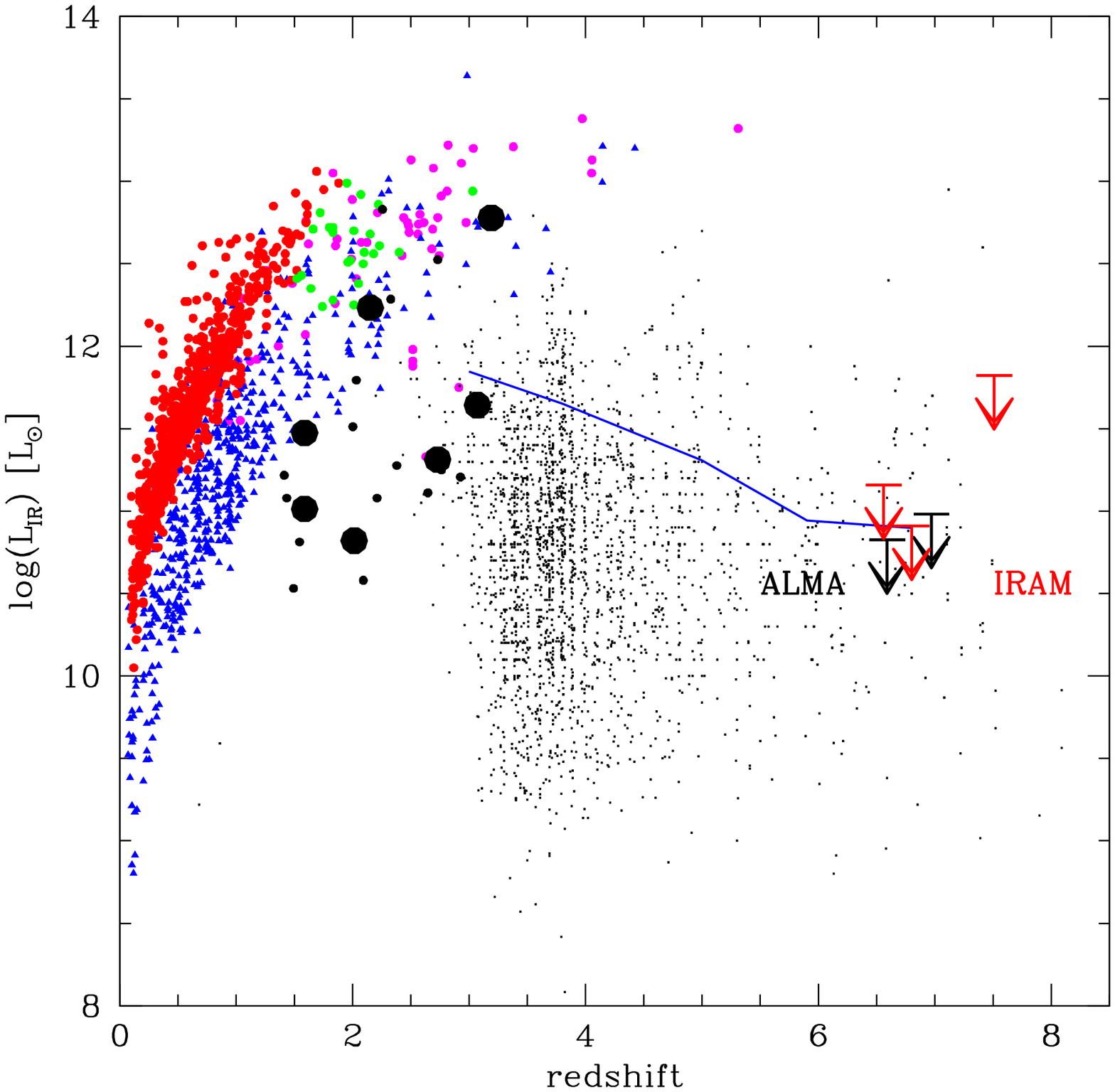}
\caption{IR luminosity (derived for $T_d=35$ K) versus redshift for the objects discussed in this paper (arrows at $z>6.5$)
and for other samples.
Colored  small circles show individual galaxies detected with  HERSCHEL in various blank fields 
(red, \citet{2013MNRAS.431.2317S}; blue \citet{Elbaz2011GOODS-Herschel:}, green: \citet{2010MNRAS.409...22M}).
The observed behavior of \lir\ with redshift for these galaxies is due to sensitivity limits of HERSCHEL,
as shown e.g.\ by  \cite{Elbaz2011GOODS-Herschel:}.
Large and small black circles show the lensed galaxies studies by \citet{Sklias2014Star-formation-}
and \citet{2013ApJ...778....2S} respectively.
The small dots show the predicted IR luminosity of LBGs from the sample studied by \cite{2014A&A...563A..81D} and
\cite{schaerer2014}. The blue line shows the IR luminosity of galaxies with a typical UV magnitude $M_{\rm UV}^\star(z)$
as predicted from the fit of \auv\ with redshift by \cite{burgarella2013}.}
\label{fig_lir_z}
\end{figure}

\section{Discussion}
\label{s_discuss}

\subsection{IR luminosity versus redshift -- comparison with other galaxy samples and LBGs}
\label{s_lir_z}

To place our upper limits on the IR luminosity in a more general context and to compare them with
IR detections of star-forming galaxies at lower redshift, we plot \lir\ versus redshift in Fig.\ \ref{fig_lir_z}.
Clearly, the current limits for the $z>6.5$ LBGs and LAE, all in the range of LIRG or sub-LIRG luminosities (i.e.\ $<10^{12}$ \lsun\ or  even $<10^{11}$ \lsun\
for some objects), are well below the detection limits of $z \ga 2$ galaxies with Herschel, but comparable to
the lowest \lir\ values of strongly lensed galaxies obtained currently at $z \sim$ 2--3.

In Fig.\ \ref{fig_lir_z} we also show the predicted IR luminosity of a sample of $\sim 2000$ LBGs between 
$z\sim 3$ and 7 analyzed by \cite{schaerer2014}. Although these values are obviously model dependent (as discussed
in \cite{schaerer2014}), this illustrates the typical range of \lir\ expected for LBGs, i.e.\ galaxies selected with
similar methods as the sources discussed here. Clearly, observations reaching similar depths as
the ones presented here, or larger depths, should soon be able to detect ``normal" LBGs over a
wide redshift range. 
For more predictions on IR luminosities of LBGs see e.g.\ \citet{schaerer2013}, \cite{2013ApJ...765....9D},
\cite{schaerer2014}, or others.
For predictions for high redshift LAEs see e.g. \cite{2009MNRAS.393.1174F,Dayal2010Detecting-Lyman}.

\begin{figure}[htb]
\centering
\includegraphics[width=8.8cm]{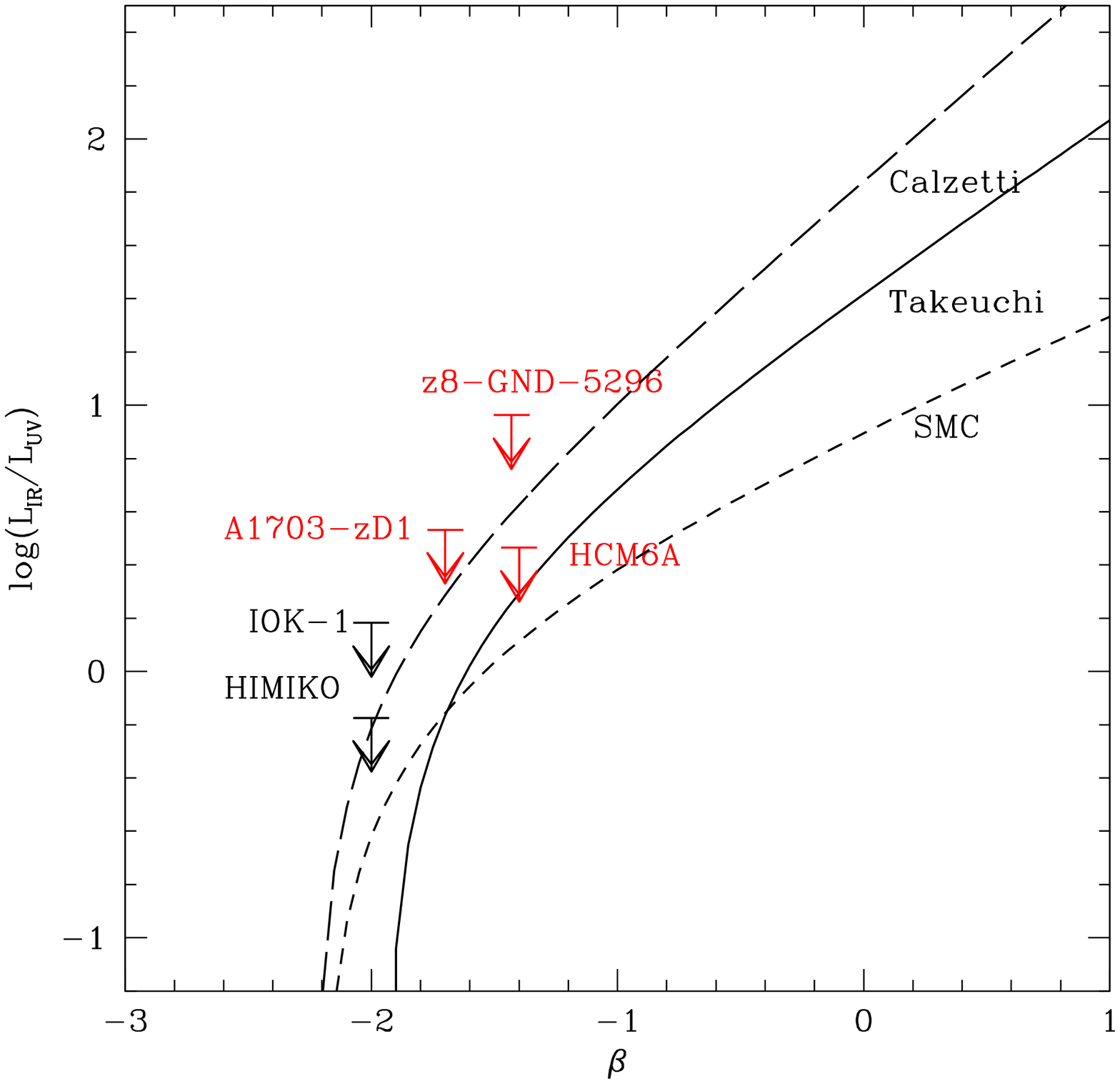}
\caption{IR/UV luminosity ratio versus UV slope $\beta$ for our objects. 
The typical uncertainty on \lir/\luv\ due to the unknown dust temperature is about $\pm 0.25$ dex;
uncertainties in $\beta$ are typically $\pm (0.3-0.5)$ (cf.\ Sect.\ \ref{s_beta}).
The long-dashed, solid, dashed lines show the relations for different attenuation/extinction laws for comparison
\citep[cf.][]{calzettietal2000,Takeuchi2012Reexamination-o}.
}
\label{fig_irx_beta}
\end{figure}

\subsection{The IRX -- beta relation}
\label{s_irx}

The ratio of the IR/UV luminosity (sometimes called IRX) is plotted as a function of the UV slope $\beta$ in Fig.\ \ref{fig_irx_beta}.
Both quantities are commonly used to determine the UV attenuation. 
Within the relatively large uncertainties,
the upper limits of \lir/\luv\ of all the $z=6.5-7.5$ galaxies studied here are compatible with expectations for normal star-forming galaxies
from the observed UV slope, 
as can be seen from comparison with the IRX--$\beta$ relations of Meurer (labelled Calzetti here) 
and the revision by Takeuchi, which describes well the locus of star-forming galaxies. Clearly, our galaxies, do not lie above 
the classical IRX--$\beta$ relation in the region where often very dusty galaxies, such as ULIRGs, are found \citep[e.g.][]{goldader02}.
The present upper limits are, however, not yet constraining enough to assess whether high-z galaxies show lower \lir/\luv\
ratios than normal star-forming galaxies. 
Note also that the two lensed galaxies observed with IRAM (\abell\ and HCM6A) provide the best constraints on the 
IRX--$\beta$ relation, comparable to the ALMA data for Himiko and IOK-1.

\begin{figure}[htb]
\centering
\includegraphics[width=8.8cm]{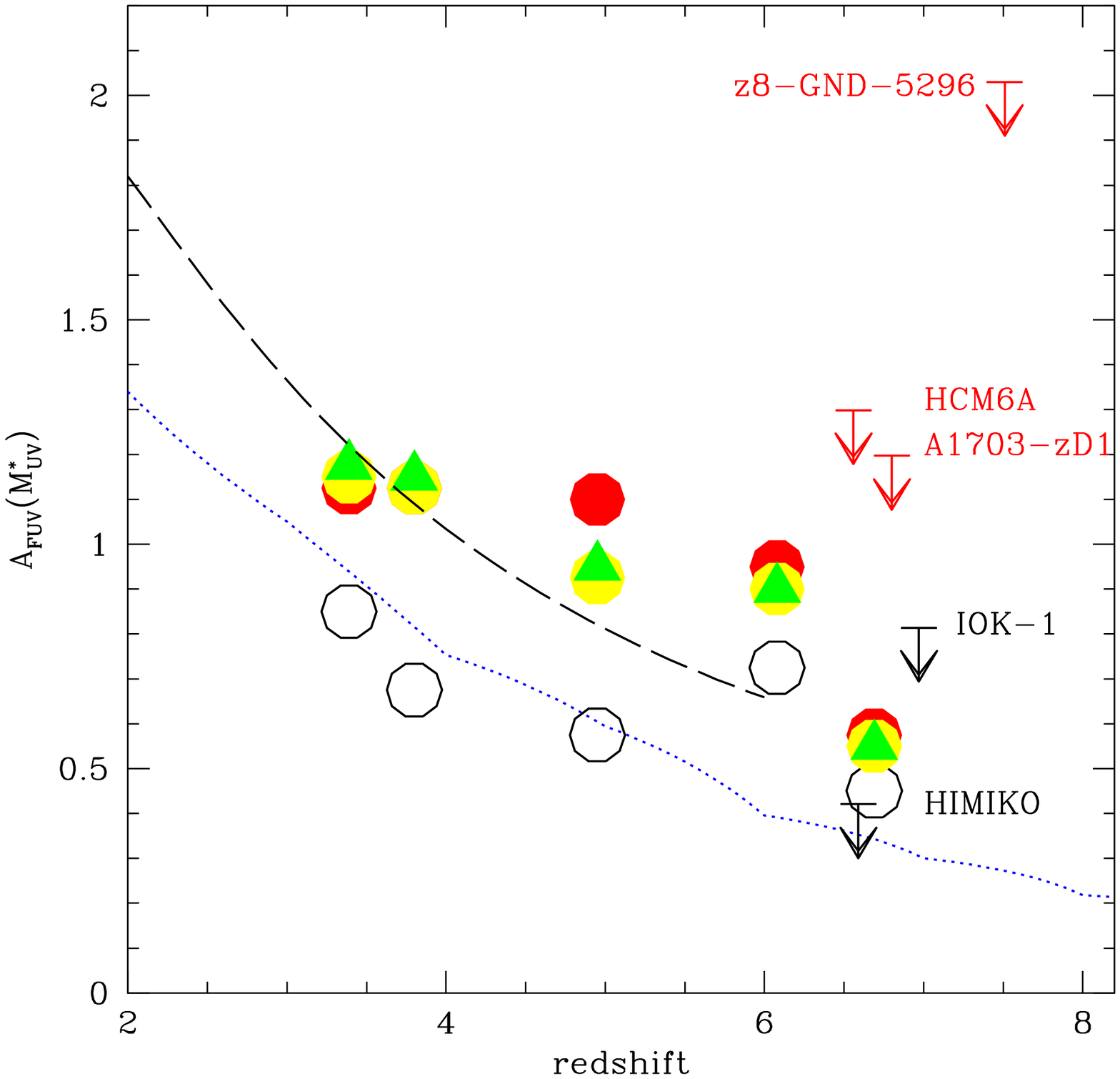}
\caption{UV attenuation \auv\ 
(upper limits, with uncertainties of $\approx 0.2-0.5$, cf.\ Table \ref{t_derived}
derived from \lir/\luv\ as a function of redshift.
The colored symbols indicate the UV attenuation at the characteristic UV magnitude $M_{\rm UV}^\star$
derived from the sample of $z \sim 3-7$ LBGs by \cite{schaerer2014} for different star-formation histories
and including nebular emission
(yellow: exponentially declining, red: exponentially rising, green: delayed SFHs).
The black circles stand for models with constant SFR neglecting nebular emission.
The long-dashed line show the extrapolation of the average UV attenuation measured from IR and UV
luminosity functions at $z \protect\la 3$ by \citet{burgarella2013}.
The blue dotted line corresponds to \protect\auv\ derived from the UV slope following 
\cite{Bouwens2013}.
}
\label{fig_av_z}
\end{figure}

\subsection{UV attenuation as a function of redshift and galaxy mass}
\label{s_uv}

The limits on the UV attenuation of the $z=6.5-7.5$ galaxies are shown as a function of redshift
in Fig.\ \ref{fig_av_z} and compared to other estimates for LBGs at these redshifts.
Again, the upper limits derived from observations are compatible with expectations
from simple relations between the UV slope and the UV attenuation (shown by the blue
dotted line)  and from our detailed SED models, which yield on average a higher
UV attenuation for the reasons already discussed by \cite{2014A&A...563A..81D} and \citet{2014A&A...566A..19C}.
The UV attenuation from these SED models are also broadly in agreement with 
the extrapolation of the cosmic attenuation with redshift proposed by
\citet{burgarella2013} from IR and UV measurements at $z \la 3$.
In short, the present data appears compatible with a median UV attenuation of 
$f_{\rm UV} \sim 1.5-2$ at $z \approx 7$, found by  \cite{schaerer2014}.
In comparison \cite{Bouwens2013} find $E(B-V)=0.02-0.03$ from the UV slope at $ z \sim 7$,
which is a factor 1.5--2 lower than the above.
%

\begin{figure}[htb]
\centering
\includegraphics[width=8.8cm]{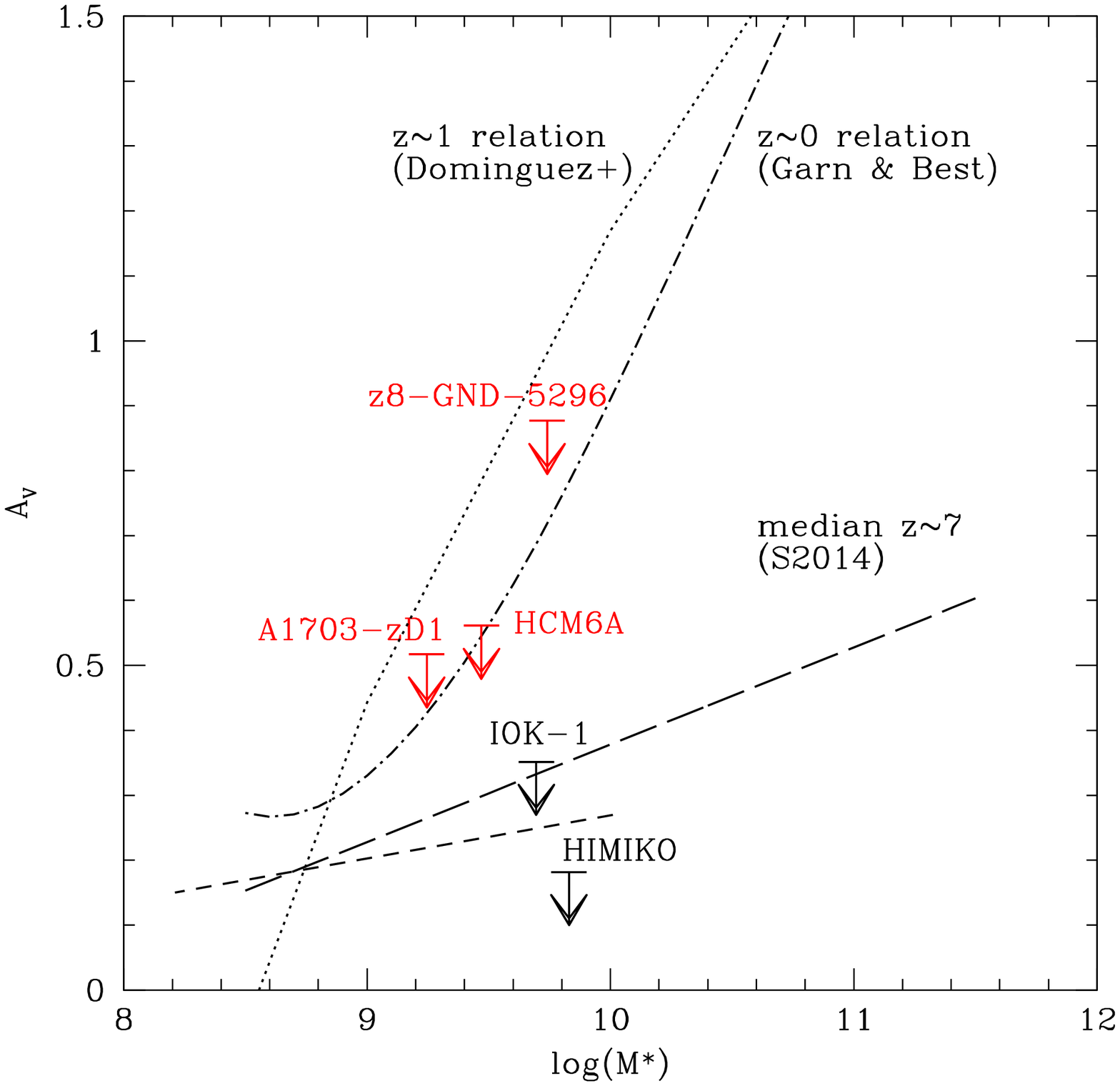}
\caption{Visual attenuation, $A_V \approx \auv /2.5$, for the objects of this study as a function of stellar mass.
The typical uncertainty on stellar mass is about $\pm 0.3$ dex 
\citep{Conroy2013Modeling-the-Pa,2014A&A...563A..81D,Sklias2014Star-formation-}.
Uncertainties on \av\ are $\approx 0.1-0.2$ (cf.\ Table \ref{t_derived}).
The dash-dotted and dotted lines show the mean relation for star-forming galaxies at $z \sim 0$ and $z \sim 0.75-1.5$
from \cite{Garn2010Predicting-dust} and \cite{dominguez2013}. 
Two median relations for LBGs at $z\sim 7$ from \cite{schaerer2014} are shown 
as long- and short-dashed lines respectively. 
The former is derived from fitting directly \mstar\ and $A_V$, the 
latter shows the median \mstar\ and $A_V$ values for LBGs with $\muv=-22$ to -$18$; both relations are for declining SFHs.}
\label{fig_mstar_av}
\end{figure}

From the nearby Universe out to $z \sim 2$ and possibly higher, various measurements
(Balmer decrement, IR/UV, and others) yield a correlation between the dust attenuation
and stellar mass, which apparently also shows little or no evolution with redshift 
\citep{Garn2010Predicting-dust,dominguez2013,2012ApJ...754L..29W,schaerer&debarros2010,
Pannella2014GOODS-HERSCHEL:}.
Even LBG samples at $z \sim 3-7$ show such a correlation 
\citep[cf.][]{schaerer&debarros2010,2014A&A...563A..81D,schaerer2014}
although these may be biased by selection effects.
It is therefore interesting to examine the constraints placed by the new UV attenuation
data as a function of the galaxy mass. This is shown in Fig.\ \ref{fig_mstar_av},
where we also plot the mean relations derived at low redshift,
the values derived from SED fits for $z \sim 4$ LBGs, and the median relation
for $z \sim 6.8$ LBGs using the same SED fitting procedure.

The upper limits for two of our galaxies (IOK-1 and Himiko) lie quite clearly below the 
local relation, indicating less dust attenuation than would be expected
on average for $z \sim 0$ galaxies with the same stellar mass. On the other hand,
the limits on dust attenuation are in good agreement or do not deviate strongly
from the median relation found from our modeling of a sample of 70 LBGs with a median
$z_{\rm phot}=6.7$ \citep{schaerer2014}. Although the upper limit for Himiko deviates most from
this relation, we do not consider this as discrepant with expectations. Indeed, 
from SED modeling one also finds relatively massive galaxies with low attenuation
\citep[cf.][]{Yabe09,2014A&A...563A..81D}, as probably also corroborated by the empirical finding of an increasing 
scatter of the UV slope towards brighter magnitudes \citep[cf.][]{2014MNRAS.440.3714R}.
%
%
In any case, the currently available samples are too small to draw significant conclusions 
on the existence or not of the mass-attenuation relation at very high redshift, and thus
on the possible redshift evolution of such a relation.

\begin{figure}[htb]
\centering
\includegraphics[width=8.8cm]{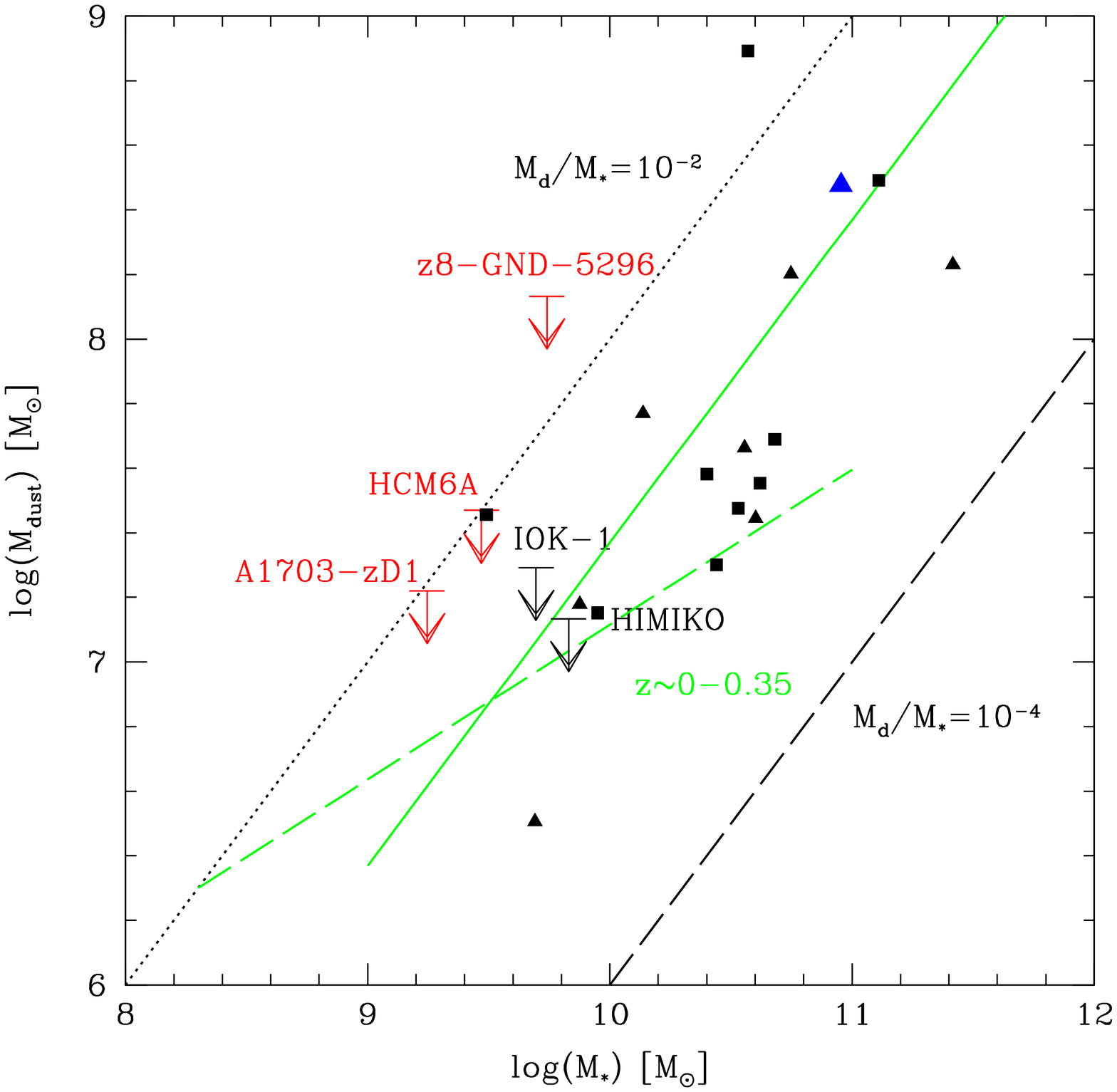}
\caption{Dust mass as a function of stellar mass for our objects (upper limits) and other related galaxies.
The typical uncertainty on \mdust\ due to the unknown dust temperature is about $\pm 0.4$ dex.
Strongly lensed galaxies at $z \sim$ 1.5--3 from \citet{2013ApJ...778....2S,Sklias2014Star-formation-}, probing 
a similar mass range, are shown as black triangles and squares respectively. 
The blue triangle shows the $z=6.34$ starbursting galaxy HFLS3 from \cite{0004-637X-790-1-40} 
with \mstar\ adjusted to the Salpeter IMF.
Typical uncertainties in stellar masses are a factor 2--3 \citep[cf.][]{Conroy2013Modeling-the-Pa,Sklias2014Star-formation-},
but can be higher in very dusty objects, such as HFLS3 \cite{0004-637X-790-1-40}.
Dotted, dashed  lines: $\mdust/\mstar=10^{-2}$, $10^{-4}$  (\mstar\ here assumes Salpeter IMF).
The green dashed line shows the location of the sequence observed by the H-ATLAS/GAMA survey at
	$z \sim$ 0--0.35 \citep{Bourne2012Herschel-ATLAS/}; the green solid line the median value of $M_d/\mstar=-2.63$ obtained
	by \cite{Smith2012} from the H-ATLAS survey after adjustment to the Salpeter IMF used here.
	The dust-to-stellar mass ratio of the high-z galaxies studied here is compatible with values observed at
	lower redshift, down to the nearby Universe. 
		}
\label{fig_mdust}
\end{figure}

\subsection{Dust mass in high-z galaxies}
\label{s_dust}

In Fig.\ \ref{fig_mdust} we plot the limits on dust mass as a function of stellar mass and compare this
to average values found in the nearby Universe and at higher redshift. We find that the present limits on dust mass are 
not incompatible with the standard dust-to-stellar mass ratios observed at low redshifts. 
{They also agree with the
recent observations of strongly lensed galaxies at $z\sim 1-3$ detected with Herschel  \citep{2013ApJ...778....2S,Sklias2014Star-formation-}.
At the high mass end, the $z=6.34$ starbursting galaxy HFLS3 detected by the HerMES survey, 
also shows a ``standard" dust-to-stellar mass ratio \citep[cf.][]{Riechers2013A-dust-obscured,0004-637X-790-1-40}. 
In short, we conclude that the available data for high redshift star-forming galaxies
may well be compatible with no significant evolution from $z \sim 0$ to 3 and out to the highest redshifts currently probed.
The current data does not show evidence for a downturn of \mdust/\mstar\ at high redshift, in contrast to
 the claim by \citet{Tan2014Dust-and-gas-in}.
As already mentioned, deeper observations and larger samples are needed to determine the 
evolution of dust with redshift, galaxy mass, and other parameters.

The ALMA observations of Himiko have been used by \cite{Hirashita2014Constraining-du}
to place a limit on the dust production per supernova in the early Universe. This limit of 0.15--0.45 \msun\
is found to be compatible with those of many nearby SN remnants.


\subsection{The \cii\ luminosity}
The \cii\ line of our IRAM targets is not detected, meaning that 
none of the five $z > 6.5$ LBGs and LAEs studied here have been detected in \cii.
Thus \abell, the three sources observed earlier, and maybe \finkel, are found to lie below the local L(\cii)--SFR relation, 
as already pointed out earlier by \cite{2013ApJ...771L..20K,Ouchi2013An-Intensely-St,Ota2014ALMA-Observatio}
for HCM6A, Himiko, and IOK-1, and as shown e.g.\ in the compilation of \cite{Ota2014ALMA-Observatio}.
The deviation from the local relation is even stronger if other SFR indicators,
such as from SED fits, are used, as these tend to yield higher star formation rates for these 
galaxies, as already mentioned above.
At somewhat lower redshift, $z=5.295$, \cite{Riechers2014ALMA-Imaging-of} have recently detected \cii\ emission
from a LBG and found it in agreement with the local L(\cii)--SFR relation. What distinguishes this object from our $z>6.5$ 
sources remains to be understood.

Possible explanations for lower \cii\ emission in high-z galaxies have already been discussed in the literature,
invoking e.g.\   compactness, lower metallicities, age effects, and others 
\citep[cf.][]{Stacey2010A-158-mum-C-II-,2013MNRAS.433.1567V,Gonzalez-Lopez2014Search-for-C-II,Ota2014ALMA-Observatio}, 
but are not fully understood.
Empirically \cii\ emission is also known to decrease in galaxies with increasing dust temperature
\citep{2011ApJ...728L...7G,2013ApJ...774...68D},  and $T_d$ has been shown to increase towards higher redshift 
\citep[e.g.][]{Sklias2014Star-formation-}.
Again, the statistics is currently too poor to draw more general conclusions on the behavior
of \cii\ emission at $z>6$ and even at lower redshifts.

\subsection{Constraints on SED models from \lir}
It is interesting to compare the limits on UV attenuation derived from the IR/UV ratio (cf.\ above) with the UV attenuation
obtained from SED fits for the same galaxies. Indeed, using \auv\ it should in principle be possible to distinguish different star formation 
histories, and direct attenuation measurements can lift some of the degeneracies
in SED fits, e.g.\ between age and attenuation, providing thus also more accurate estimates of the SFR, stellar mass, hence also sSFR
\citep[cf.][]{reddy2012,schaerer2013,Sklias2014Star-formation-}.

\subsubsection{Constraints on SF histories?}
Comparing the results from our SED fits where extinction is kept as free parameter 
with the \auv\ values
from the IR/UV ratio 
we find useful constraints for two objects, \abell\ and Himiko.
SED fits assuming exponentially rising or declining star formation histories (SFHs) yield quite young ages ($\sim 10-40$ Myr) and a high extinction
$A_V \sim 0.5-0.95$, i.e.\ $\auv \sim 1.1-2.2$,  (quoting here the range covered within 68 \% confidence for the Calzetti law), for 
these objects and also for \finkel.  \citet{Ouchi2013An-Intensely-St} find values comparable to ours for Himiko,
translating to $\auv = 1.38$ with an uncertainty less than 0.1.
Within 68\% confidence this high attenuation is excluded by the ALMA observations of Himiko. 
With rising SFHs we always find  tight uncertainties on the attenuation for this object, the lowest value
being $\auv=1.0$ within 99\% confidence. For constant SFR or declining SFHs, however, we obtain 
lower values for the attenuation, compatible with the limit $\auv < 0.4^{+0.3}_{-0.2}$ from ALMA within 90--95\% confidence.
Taken at face value this implies that  exponentially rising star-formation histories are not a good description
for the stellar populations of Himiko, as this would predict too high a UV attenuation. 
For \abell\ the trend is similar, although the less strong limits on \auv\ imply that the SED models are
still in agreement within $\ga$ 68-90 \% confidence.

We then ran SED fitting models where the UV attenuation is limited to the value derived above, i.e.\ \auv\ from 
Table \ref{t_derived}. We now discuss how this may affect the physical parameters, among which in 
particular the specific star formation rate.
 
\begin{figure}[htb]
\centering
\includegraphics[width=8.8cm]{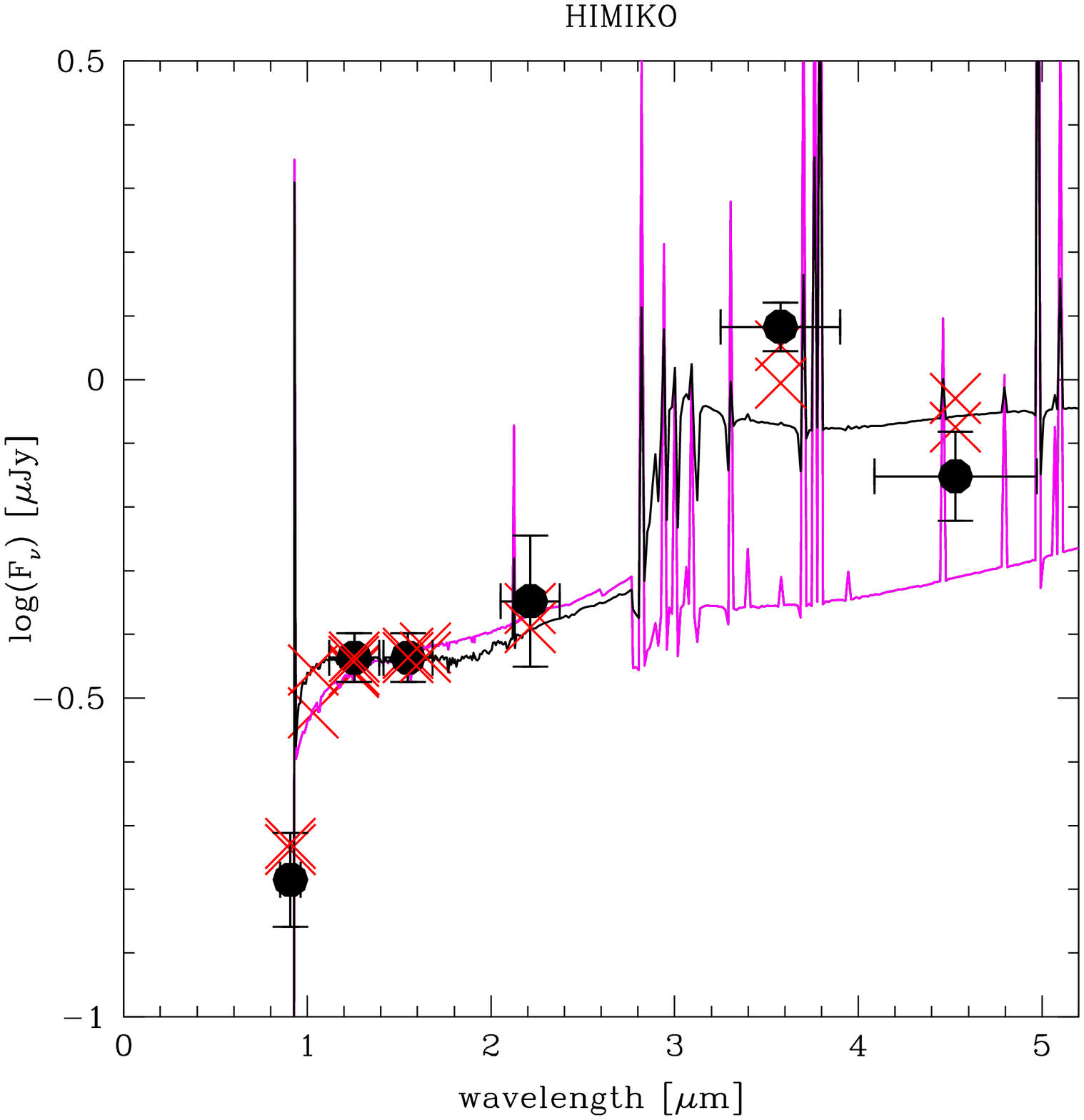}
\caption{Best-fit SEDs for Himiko with unconstrained models (magenta) or imposing the constraint on
UV attenuation of $\auv<0.4$ (black). Red crosses show the synthesized fluxes in the photometric bands;
black circles the observations.
The two solutions differ strongly in age, extinction, SFR, mass and sSFR. The young one (magenta) 
overproduces the IR luminosity (see text).
The IRAC channel 1 (3.6 \micron) is contaminated by \Oiii\ and \hb\ emission lines; \ha\ contributes
weakly to channel 2 (4.5 \micron). The relatively weak emission line at $\sim$ 4.5 \protect\micron\ is He~{\sc i} $\lambda$5876.
}
\label{fig_Himiko}
\end{figure}
\begin{figure}[htb]
\centering
\includegraphics[width=8.8cm]{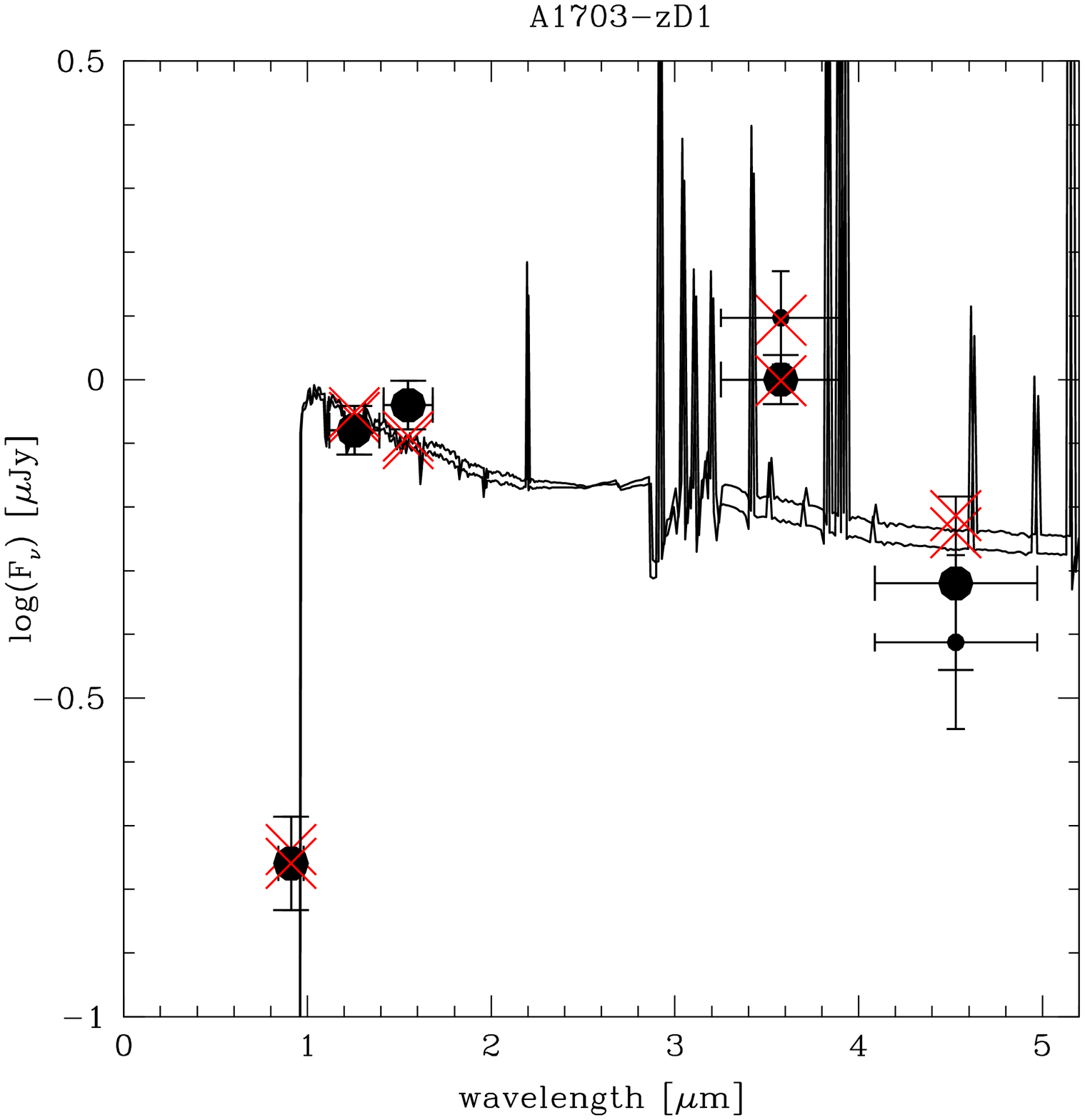}
\caption{Best-fit SEDs for \abell\ with constrained models ($\protect\auv <1.2$)  to two different sets of IRAC measurements 
(cf.\ Sect.\ \ref{s_otherobs}). Red crosses show the synthesized fluxes in the photometric bands;
black circles the observations.
In both cases the resulting fit parameters are quite similar confirming a high specific SFR for this galaxy. The same
is also obtained without the constraint on UV attenuation.}
\label{fig_abell}
\end{figure}

\subsubsection{High specific SFR at $z \sim 6.5-7.5$?}
We first examine Himiko, with the tightest limit on UV attenuation, by imposing $\auv <0.4$ (i.e.\ $A_V <0.17$) on
SED fits. The resulting best-fit SED for models with declining SFHs is shown in Fig.\ \ref{fig_Himiko}, where it is 
also compared to the unconstrained SED fit yielding $A_V=1$. As clearly seen, the two fits differ strongly in 
a qualitative way, the former representing a relatively old population with a low attenuation, whereas the latter
case is much younger, has stronger emission lines and a higher attenuation. Correspondingly the older population
yields a lower SFR and higher stellar mass, hence a lower sSFR than the unconstrained fits with
\mstar\ changing from $\sim 1.0 \times 10^{10}$ to $4.2\times 10^{10}$ \msun\ , SFR from $\sim$ 160 to 20 \msunyr, and the specific SFR by a factor $\la$ 40.
For exponentially rising SFHs with the \auv\ constraint, we obtain  $A_V=0.15$, $\mstar \sim 2.0 \times 10^{10}$  \msun\ , SFR $\sim 30$ \msunyr, 
i.e.\ sSFR $\sim 1.5$ Gyr$^{-1}$.
It is therefore possible that Himiko has a relatively low sSFR (even close to the often quoted, ```conservative" value of 1--2 Gyr$^{-1}$
\citep[cf.][]{2014ApJ...781...34G}). 
Part of the difficulty to distinguish the above solutions stems from the fact that both IRAC filters
can be contaminated by emission lines at the redshift of Himiko ($z=6.595$), and from the finding 
of the 4.5 \micron\ flux observed stronger than at 2.2 \micron.
Measurements of the rest-frame optical emission lines of \Oiii, \hb\ (in the 
3.6 \micron\ filter) and/or \ha\ at 4.5 \micron, or an accurate determination of the continuum flux at these wavelengths
will be necessary to precisely determine the current SFR of this galaxy and its stellar mass.

For \abell\ the situation seems clearer, since the observed 4.5 \micron\ flux is weak, fainter than at
near-IR (1--2 \micron), and a significant excess at 3.6 \micron\ is found compared to 4.5 \micron\
(see Fig.\ \ref{fig_abell})\footnote{The interpretation of the SED of \abell\ is also clearer since its
redshift is most likely $z \sim 6.8 \pm 0.1$, in a range where \ha\ has moved out of the 4.5 \micron\ filter 
providing thus a large contrast between the 3.6 and 4.5 \micron\ bands, as also highlighted by \cite{2014ApJ...784...58S}.}
In this case an old population (significant Balmer break) is clearly excluded,
i.e.\ the stellar mass must be relatively low, and the (strong) 3.6 \micron\ excess implies a high current
SFR. Hence the specific SFR of this galaxy is much better constrained than for Himiko, and 
must be fairly high, as also found by \cite{2014ApJ...784...58S}.
For \abell, models both with or without imposed constraints on the UV attenuation, 
with declining or rising SFHs, and for two different measurements of the IRAC photometry, 
yield indeed a high sSFR $\sim$ 20--90 Gyr$^{-1}$.

In short, solutions with very young ages, hence high SFR, lower mass, and the highest sSFR are disfavored
for some of our galaxies as Himiko.
The young population \citet{2013Natur.502..524F} and we find for \finkel\ is not excluded by the current IRAM data,
leaving, however, large uncertainties on the physical parameters of this object.
To firm up conclusions on the sSFR of HCM6A and IOK-1, accurate IRAC flux measurements are needed
(cf.\ Sect. \ref{s_otherobs}). In any case, since both IRAC filters are affected by nebular emission lines at the redshift of these 
galaxies, degeneracies will remain between the derived physical parameters. Direct emission line measurements
(e.g.\ with the JWST) will be crucial to firmly establish the properties of such high redshift galaxies.


\section{Conclusion}
\label{s_conclude}
We have carried out deep 1.2mm observations with the Plateau de Bure Interferometer, targeting [CII] $\lambda$158\micron\ and dust continuum emission
of  two $z \sim 7$ Lyman break galaxies, \abell\ and \finkel\ discovered earlier by \cite{Bradley2012Through-the-Loo} and  \citet{2013Natur.502..524F}.
\abell\ is a bright LBG, magnified by a factor $\mu=9$ by the lensing cluster Abell 1703. \finkel\ is currently one of the most distant, spectroscopically
confirmed galaxies at $z=7.508$.

We have combined our observations with those of other  $z>6.5$ star-forming galaxies with spectroscopic redshifts (three LAEs named HCM6A, Himiko, and IOK-1),
for which deep measurements were recently obtained with the PdBI and ALMA  \citep{2013ApJ...771L..20K,Ouchi2013An-Intensely-St,Ota2014ALMA-Observatio}.
For this small sample of five high-$z$ star-forming galaxies we have determined in a homogenous manner their 
IR hence dust properties, as well as their stellar emission, from which we derive limits on the IR luminosities and star formation rates, dust masses, 
the \cii\ luminosity, UV attenuation, and stellar masses.

Our main results are the following:
\begin{itemize}
\item The dust continuum of \abell\ and \finkel\ is not detected down to an rms of 0.12 and 0.16 mJy/beam at 241 and 223 GHz, respectively.
From this we obtain upper limits on the IR luminosity of $\lir(T_d=35)<8.1 \times 10^{10}$ \lsun\ and $<6.7 \times 10^{11}$ \lsun\
for  \abell\ and \finkel, assuming a dust temperature $T_d=35$ K prior to correction for CMB heating. 
Thanks to strong gravitational lensing, our observations of \abell\ reach a similar effective depth as 
the recent ALMA observations of Himiko and IOK-1, which also probe the sub-LIRG ($\lir < 10^{11}$ \lsun) regime.

\item The upper limits of \lir/\luv\ of the five galaxies and their observed UV slope $\beta$ follow quite closely the 
so-called IRX--$\beta$ relation expected for the Calzetti attenuation law.

\item The UV attenuation, derived from the ratio of \lir/\luv, is found to be between $\auv < 0.4^{+0.3}_{-0.2}$ for Himiko
(the lowest limit), and $\auv < 1.2-1.3$ for the lensed galaxies \abell\ and HCM6A.
Broadly, these limits are compatible with extrapolations of \auv\ measurements from the IR and UV at lower redshift
\citep{burgarella2013} and with classical estimates from the UV slope \citep[e.g.][]{Bouwens2013}.
The available limits are also compatible with results from recent SED fits for large samples of LBGs, which
indicate that the UV attenuation may be higher than derived using the standard relation between the UV slope
and attenuation \citep{2014A&A...563A..81D,2014A&A...566A..19C}.

\item For their stellar mass (of the order of $\mstar \sim  (2-10) \times 10^9$ \msun) the high-$z$ galaxies
studied here have an attenuation below the one expected from the mean relation observed for local galaxies
and out to $z \sim 1.5$ \citep[cf.][]{dominguez2013}.
The current limits are, however, compatible with the median relation between and attenuation derived
by \cite{schaerer2014} for $z \sim 7$ LBGs.

\item  The limits on the dust masses for  \abell\ and \finkel\ are $M_d(T_{\rm d}=35) <1.6 \times 10^7$ \msun\ and
$<13.9  \times 10^7$ \msun\ respectively. The available data/limits for $z>6$ galaxies show 
dust-to-stellar mass ratios, which are not incompatible with that of galaxies from $z=0$ to 3. 
Deeper observations are required to establish whether the dust-to-stellar mass ratio decreases at high redshift,
e.g.\ due to lower metallicity.

\item If the redshift of \abell\ is confirmed within our targeted range ($z \in [6.796,6.912]$) our non-detection
for this galaxy adds another limit on the \cii\ $\lambda$158\micron\ luminosity ($\lcii <2.8 \times 10^7$ \lsun), showing that the $z>6$
galaxies observed so far emit less \cii\ than expected from their SFR when compared to low redshift galaxies
\citep[cf.][]{Ouchi2013An-Intensely-St,Ota2014ALMA-Observatio}.
\cii\ emission from \finkel\ is not detected with an upper limit of $\lcii <3.6 \times 10^8$ \lsun.

\item The derived limits on the UV attenuation exclude some SEDs fits suggesting very young ages and
may place better limits on the specific SFR at high redshift. However, the presence of emission lines in several
IRAC filters currently prevents accurate determinations of the SFR and stellar mass of $z>6$ galaxies,
except over a narrow redshift interval \citep[$z \sim 6.8-7.0$; cf.][]{2014ApJ...784...58S}.
For \abell\ we confirm the very high sSFR $\sim$ 20--90 Gyr$^{-1}$ of \abell\ derived by \citet{2014ApJ...784...58S}.
\end{itemize}

Although our new measurements provide some additional insight, more and deeper (sub)-mm data, 
both of lensed and blank field galaxies, are clearly needed to determine the UV 
attenuation and dust content of the dominant population of the most distant star-forming galaxies
and their evolution with redshift, and to understand the apparent lack of \cii\ emission from these
galaxies and their ISM properties more in general.

\begin{acknowledgements}
We thank IRAM, in particular Tessel van der Laan, for efficient observations and help with data reduction.
This work was supported by the Swiss National Science Foundation.
J.S. and the GISMO observations were supported through NSF ATI grants 1020981 and 1106284.
We made use of the public {\em Cosmolopy} python package from Roban Kramer ({\tt http://roban.github.com/CosmoloPy/}),
the python version of Ned Wright's cosmology calculator, from James Schombert,
TOPCAT \citep{2005ASPC..347...29T}, and the NASA ADS services.
This paper makes use of the following ALMA data:
ADS/JAO.ALMA\#2011.0.00115.S and ADS/JAO.ALMA\#2011.0.00767.S. 
ALMA is a partnership of ESO (representing
its member states), NSF (USA) and NINS (Japan), together with NRC
(Canada) and NSC and ASIAA (Taiwan), in cooperation with the Republic of
Chile. The Joint ALMA Observatory is operated by ESO, AUI/NRAO and NAOJ.
\end{acknowledgements}



\bibliographystyle{aa}
\bibliography{merge_misc_highz_literature}

\end{document}